\documentclass{jaa}
\usepackage{natbib}
\usepackage{graphicx}
\usepackage{hyperref}
\usepackage{color,colortbl}
\usepackage{tablefootnote}
\definecolor{Gray}{gray}{0.9}
\definecolor{LightCyan}{rgb}{0.88,1,1}

\newcommand{\ps}{}

\begin{document}\sloppy

\title{
Computational Astrophysics, Data Science \& AI/ML in  Astronomy: A Perspective from Indian Community}


\author{Prateek Sharma\textsuperscript{1,*}, Bhargav Vaidya\textsuperscript{2,*}, Yogesh Wadadekar\textsuperscript{3,*}, Jasjeet Bagla\textsuperscript{3, 4}, Piyali Chatterjee\textsuperscript{5}, Shravan Hanasoge\textsuperscript{6}, Prayush Kumar\textsuperscript{7} Dipanjan Mukherjee\textsuperscript{8}, Ninan Sajeeth Philip\textsuperscript{9}, Nishant Singh\textsuperscript{8}}
\affilOne{\textsuperscript{1}Department of Physics, Indian Institute of Science, Bangalore 560012, India\\}
\affilTwo{\textsuperscript{2}DAASE, Indian Institute of Technology, Indore, Madhya Pradesh, 453552, India\\}
\affilThree{\textsuperscript{3}NCRA, Tata Institute of Fundamental Research, Ganeshkhind, Pune 411007, India\\}
\affilFour{\textsuperscript{4}IISER Mohali, Knowledge City, Sector 81, SAS Nagar, Punjab 140306, India\\}
\affilFive{\textsuperscript{5}Indian Institute of Astrophysics, II Block Koramangala, Bengaluru-560034, India\\}
\affilSix{\textsuperscript{6}Department of Astronomy and Astrophysics, Tata Institute of Fundamental Research, Mumbai, India\\}
\affilSeven{\textsuperscript{7}ICTS, Tata Institute of Fundamental Research, Bengaluru 560089, India\\}
\affilEight{\textsuperscript{8}Inter-University Centre for Astronomy \& Astrophysics, Post Bag 4, Ganeshkhind, Pune 411007, India\\}
\affilNine{\textsuperscript{9}Artificial intelligence Research and Intelligent Systems, Thadiyoor-Valakuzhy Rd, Valakuzhy, Kerala 689544, India}


\twocolumn[{

\maketitle

\corres{prateek@iisc.ac.in, bvaidya@iiti.ac.in, yogesh@ncra.tifr.res.in}



\begin{abstract}
\ps{In contemporary astronomy and astrophysics (A\&A), the integration of high-performance computing (HPC), big data analytics, and artificial intelligence/machine learning (AI/ML) has become essential for advancing research across a wide range of scientific domains. These tools are playing an increasingly pivotal role in accelerating discoveries, simulating complex astrophysical phenomena, and analyzing vast amounts of observational data. For India to maintain and enhance its competitive edge in the global landscape of computational astrophysics and data science, it is crucial for the Indian A\&A community to fully embrace these transformative technologies. Despite limited resources, the expanding Indian community has already made significant scientific contributions. However, to remain globally competitive in the coming years, it is vital to establish a robust national framework that provides researchers with reliable access to state-of-the-art computational resources. This system should involve the regular solicitation of computational proposals, which can be assessed by domain experts and HPC specialists, ensuring that high-impact research receives the necessary support. By building such a system, India can cultivate the talent, infrastructure, and collaborative environment necessary to foster world-class research in computational astrophysics and data science.}
\end{abstract}

}]

\pgrange{1--}
\setcounter{page}{1}
\lp{21}

\section{Introduction}
High Performance Computing (HPC) – the use of a large number of processors in parallel to solve big computational problems –  is increasingly playing an important role in astrophysics research; e.g., gravitational N-body dynamics, (magneto)hydrodynamics of galaxy formation, accretion and jet physics around compact objects, and dynamics of spacetime geometry due to compact object mergers. The aims of these simulations range from understanding the basic physical phenomena (e.g., the growth and saturation of the magnetorotational instability) to producing mock observables using subgrid physics that can be compared directly with observations (e.g., mock Event Horizon Telescope images from MHD simulations of accretion around black holes using subgrid prescriptions for electron temperature/heating from MHD fields, and mock gravitational-wave signals from numerical relativistic simulations of black hole binary mergers). Similarly, the data gathered by telescopes and generated by massive simulations have grown exponentially and need to be carefully analyzed and organized for their effective future use. Artificial intelligence (AI) and Machine Learning (ML) techniques – which are heavily influencing our daily lives – are also increasingly used to 
quickly discover features in astrophysical images, lightcurves and spectra, and characterise, cluster, and classify sources and populations.

The Indian astronomy community must stay at the forefront of these exciting developments in astrophysical applications of HPC, data and AI/ML (artificial intelligence/machine learning). For this, we need to constantly adapt our teaching/training of the students/workforce and make available to them the required hardware and software resources. An urgent need is for a framework to provide quick and fair access to HPC resources in the country to deserving astrophysics projects. While the National Supercomputing Mission (NSM) has created a handful of PetaFlop clusters across India, access to these is not through regular calls for proposals. We need to quickly evolve a process for the allocation of national HPC resources through periodic (say, four times a year) calls for proposals that are evaluated by programme committees from different research areas that use HPC. This is close to the procedure followed in more mature HPC ecosystems such as in the USA and Europe. 

This chapter starts with the global status of HPC/data/AI/ML in astrophysics, presents a glimpse of research in this area by the Indian astronomy community, reviews key questions and aims at future goals, and makes a case for observational/computational facilities. We finish by providing short and medium/long-term goals in astrophysical HPC/data/AI/ML and the broader applicability of these techniques.

\section{Global status}

In this section we briefly survey global trends in HPC, some prominent astrophysical applications of HPC, and the increasing of AI/ML techniques in astronomy.

\subsection{HPC: Access \& Hardware/Software Landscape}

On the world stage, there are several national projects that provide researchers access to state-of-the-art HPC facilities for conducting their computational and data-intensive research. Some of these projects are listed below.
\begin{itemize}
\item ACCESS\footnote{\url{https://access-ci.org/about/}} is the HPC project (funded by NSF) that researchers in the US can use to access supercomputing resources through computing allocations based on scientific proposals. It succeeds similar frameworks in the past: TeraGrid and XSEDE. 
\item NASA’s High-End Computing Capability (HECC)\footnote{\url{https://www.nas.nasa.gov/hecc/}} also provides access to supercomputers like Pleiades and Aitken based on scientific proposals, or if the user is already a part of any of its space/ground-based mission teams. The US space agency provides computing resources as part of all its space and astronomy missions to augment the science that comes out of these expensive instrumentation endeavors. 
\item PRACE\footnote{\url{https://prace-ri.eu/about/introduction/}} is the framework that provides HPC infrastructure to European researchers.
\item In addition to the ones listed above, similar infrastructure projects in other countries such as Japan\footnote{\url{https://fugaku100kei.jp/e/exhibit/}}, Germany\footnote{\url{https://www.mpcdf.mpg.de}} and Australia\footnote{\url{https://discover.pawsey.org.au/project/setonix}}, provide tiered access to HPC resources across a range of disciplines based on proposals.
\end{itemize}

Most of the supercomputing resources are primarily available to only resident researchers. These policies adopted in mature supercomputing nations, based on tiered access and {\em quick proposal-based evaluations to access HPC resources} for both academic research and industries, are urgently needed in India. Moreover, access to supercomputers should depend solely on the merit of proposals rather than the physical location of the supercomputer. It is easy to achieve high utilisation of HPC resources without a significant scientific return or societal value in absence of a quick, transparent, and reasonable evaluation framework. 

Access to advanced cyber-infrastructure leads to {\em measurable increase in the impact of scientific research} (e.g., \citealt{Wang2018}). Publications across different fields using XSEDE infrastructure were found to receive a factor of 3 to 7 times larger citations compared to the average citations in the field, showing that publications with higher impact result from access to advanced HPC resources across all fields. Moreover, several of XSEDE-supported papers are among the highly cited papers in their fields.

\begin{figure}
    \centering
    \includegraphics[width=1\columnwidth]{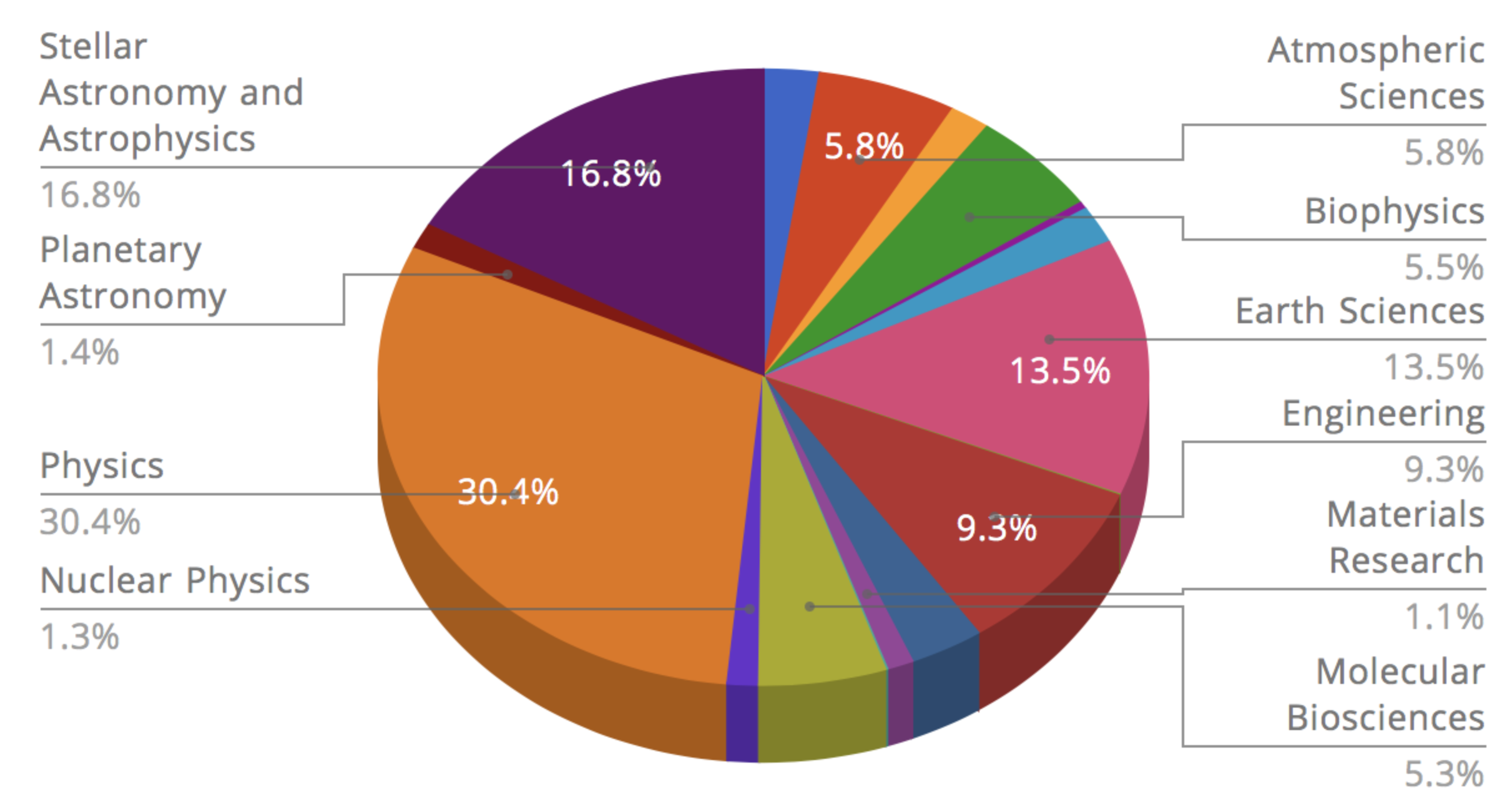}
    \caption{A pie chart depicting the division of workload at Blue Waters among different scientific disciplines as of May 2017. The chart is adapted from \url{https://www.hpcwire.com/2017/05/02/blue-waters-study-dives-deep-performance-details/}}
    \label{fig:pie_hpc_workload}
\end{figure}

The June 2022 distribution of application areas of top 500 clusters\footnote{\url{https://top500.org}} shows that about 26\% of the system share is devoted to research. Smaller fractions are devoted to weather \& climate (9\%), energy (9\%), aerospace (4\%), and semiconductor (4\%) research. While most of the systems are in the industry segment (47\%), research and academic organisations account for about 40\% of the system share. 
An analysis of workload for the Blue Waters\footnote{\url{https://bluewaters.ncsa.illinois.edu}} from 2013 to 2017 shows that about 17.3 billion core hours had been dedicated to scientific research. A pie chart showing the division among different scientific disciplines is shown in figure~\ref{fig:pie_hpc_workload}. It is evident that close to 20\% of the available research computing is taken up by stellar, planetary and astrophysics research. 

Turning to the most recent data on usage of US supercomputing resources across different fields, ACCESS data\footnote{\url{https://xdmod.access-ci.org}} shows that about 192 million CPU hours were used by Astronomy \& Planetary Sciences on ACCESS supercomputers for the year 2022, only behind Materials Engineering that used 265 million CPU hours, and exceeding all the other listed 74 research fields. ACCESS data also shows that over time HPC resources are used in more and more scientific areas, indicating the increased importance of HPC across different fields.

 These data, taken as a proxy (albeit imperfect) for the extent of HPC usage in various academic disciplines, illustrate that {\em Astronomy and Astrophysics is one of the top users of HPC resources worldwide}. A similar analysis over 2013-2016 on Blue Waters supercomputer, dedicated to academic research in science and engineering, showed that the majority of node hours were consumed by Physical and Astronomical Sciences (\citealt{Jones2017}). Moreover, the {\em majority of the node hours were  occupied by well-tested community codes} (rather than individual groups’ codes) in different fields. 
 Such community codes in astrophysics include PLUTO, Pencil Code, Athena, RAMSES, Gadget, Enzo, SpEC, ET and SpECTRE (c. f. Table \ref{tab:codes_india}).

The latest data from the top 500 cluster shows that India needs to invest heavily in HPC in order to catch up with the rest of the world in this important area. The national supercomputing mission (NSM), under which HPC infrastructure is being set up in several research institutes in India, is a step in that direction. As the experience of other countries shows, efficient management and access to computing based on strong scientific proposals are key to maximising returns on the investment in HPC. 

The Exascale Computing Project (ECP) is the US Department of Energy’s (DOE) \$4 billion 7-year program to maintain the US's competitive edge in HPC, with a focus on developing the exascale ecosystem (R\&D, hardware/software/HR development, developing exascale-ready applications from across scientific/strategic areas including astrophysics) and installing 3 exascale machines in DOE labs\footnote{\url{https://www.exascaleproject.org/about/}}. It is pretty clear that efficiency will be achieved by writing software that is aware of hardware and new algorithms, rather than simply relying on faster multicore processors via Moore’s law (\citealt{Leiserson2020}). This new HPC landscape requires specialized knowledge of computing and close coordination between hardware, software, and applications teams.

A major trend in recent top 500 cluster data is the {\em increasing share of GPU (graphical processing unit) accelerators/coprocessors} in the largest supercomputers. The heterogeneous architectures also fare better on energy efficiency. To efficiently utilize the GPUs on Petascale and Exascale systems, several popular astrophysics codes such as Athena, PLUTO and Enzo are developing GPU versions.

\subsection{Computational Astrophysics Research} 

\label{comp_global}

On a global scale, the supercomputing facilities mentioned earlier have played an important role in enabling cutting-edge simulations in Astronomy and Astrophysics. The ultimate goal of such simulations is to generate synthetic data that can be compared to observations. In the last decade, significant progress has been made in both  observations and simulations. Large multi-wavelength surveys and follow-ups of individual objects have uncovered the statistics and details of various astrophysical phenomena. Understanding such observations necessitates advanced numerical simulations that can ``reproduce" them. The global computational astrophysics community has made significant progress in this direction.

\begin{figure*}
\centering
\includegraphics[width=2\columnwidth]{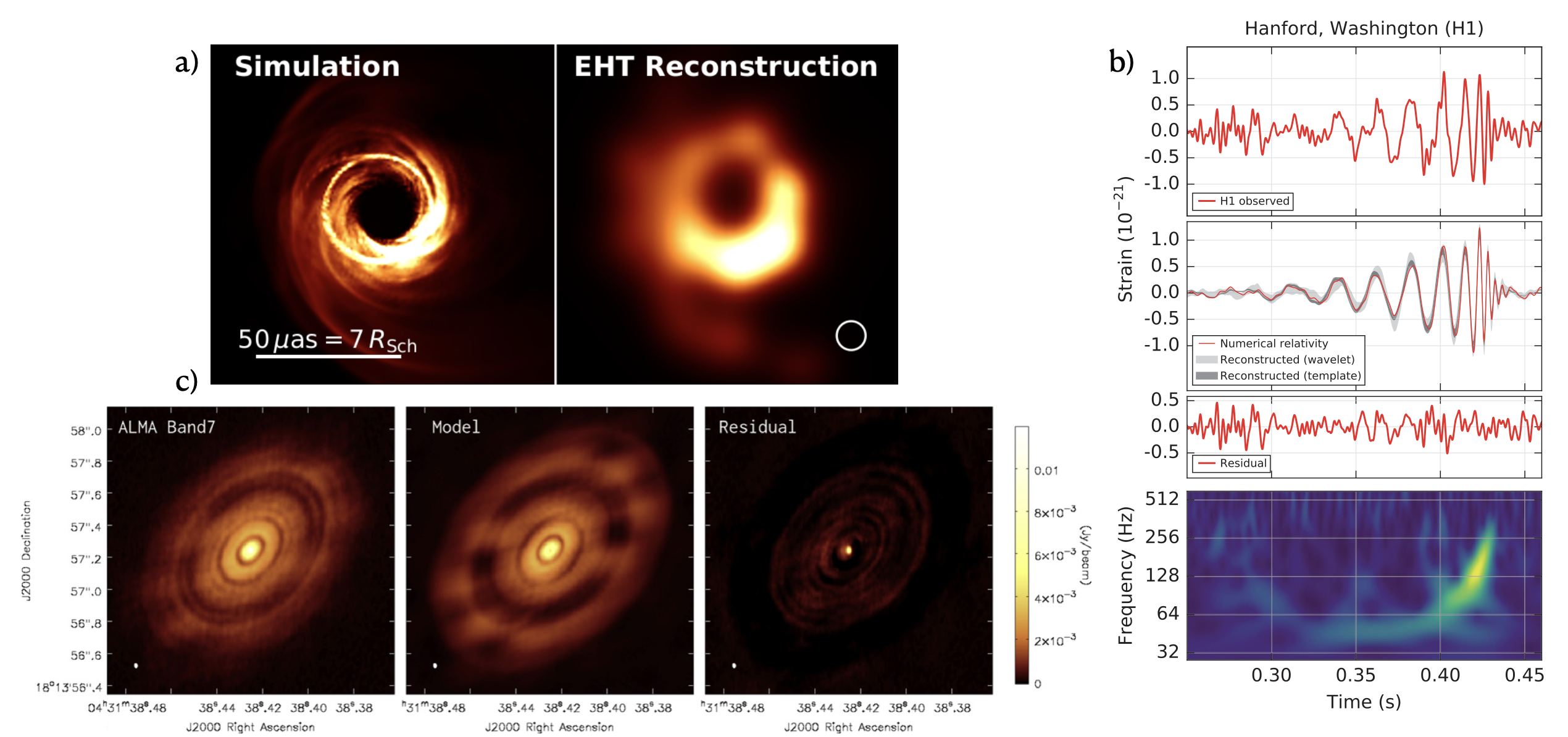}
\caption{(\textit{Panel a})- Comparing simulations with the reconstructed image of M87 black hole obtained from EHT (adapted from \citealt{Chael2021})
(\textit{Panel b}) - The gravitational-wave event GW150914 observed by the LIGO Hanford (adapted from \citealt{LIGO2016}), where the waveform from numerical relativity model (red line) is shown in comparison with the observed strain in the middle panel.
(\textit{Panel c}) - Synthetic image of HL Tau disk from the model compared to ALMA Band 7 data and the residual image (adapted from \citealt{Jin2016}).}
\label{fig:gl_compastro}
\end{figure*}

Major bottleneck confronting 
computational astrophysics is the range of physical scales to be simulated in both space and time. Furthermore, incorporating the complex interplay of several physical processes (e.g., radiative transfer, chemistry, plasma transport) that produce emission in these sources complicates the computation. Despite these odds, there have been successful attempts worldwide to replicate some of the observations from recent discoveries. Here we list some prominent examples:

\begin{itemize}
\item Full three-dimensional general relativistic magnetohydrodynamic (GRMHD) simulations of the accretion flow around spinning black holes have been performed, and the images observed with the Event Horizon Telescope for the Kerr black hole at the center of M87 have been successfully reproduced with subgrid physics (e.g., \citealt{Chael2021}). Pipelines have also been developed to simulate the electromagnetic observables of black hole accretion, such as PATOKA (\citealt{Wong2022}), allowing simulation data to be converted into synthetic observations (see panel a of figure \ref{fig:gl_compastro}).
\item Numerical simulations combined with post-processing radiative transfer codes such as RADMC-3D were successful in reproducing the emission signatures of the dusty proto-planetary disc HL Tau, including the gaps observed in the ALMA band (see panel c of figure ~\ref{fig:gl_compastro} adapted from \citealt{Jin2016}))
\item The IllustrisTNG project\footnote{\url{https://www.tng-project.org/about/}} is a collection of cutting-edge simulations of cosmological galaxy formation. Each IllustrisTNG simulation evolves a large swath of a mock Universe from shortly after the Big Bang to the present day while accounting for a wide range of physical processes – gravity, gas dynamics, cooling, and feedback heating – that drive galaxy formation. These simulation data are shared over the web and are being used across the world to investigate a wide range of questions concerning the evolution of the Universe and its galaxies over time (e.g., \citealt{Pillepich2018,Nelson2019}). 
\item Fully general relativistic simulations of black hole binary systems (\citealt{LIGO2016}), and fully dynamical GRMHD simulations of neutron star binaries (\citealt{Shibata2017}) have been key to extracting information from gravitational-wave signals observed by the LIGO \& Virgo detectors\footnote{\url{https://www.ligo.org/}} (soon to be joined by LIGO-India\footnote{\url{https://www.ligo-india.in/}}). These are carried out by international collaborations such as the SXS project\footnote{\url{https://www.black-holes.org/}} that aim to simulate black holes and other extreme spacetimes in order to gain a better understanding of relativity and the physics of exotic objects in the distant cosmos. Their flagship Spectral Einstein Code (SpEC\footnote{\url{https://www.black-holes.org/code/SpEC.html}}) has been instrumental in following up GW detections so far (e.g., see right panel of Figure~\ref{fig:gl_compastro}). Their next-generation open-source SpECTRE code\footnote{\url{https://spectre-code.org/}} (\citealt{Deppe2022}) can numerically solve multi-scale, multi-physics problems in astrophysics and numerical relativity using petascale and exascale computing, and endeavors to gradually generalize to problems across discipline boundaries in fluid dynamics, geoscience, plasma physics, nuclear physics, and engineering.

\item Now we give an example of solar convection where numerical models
being pursued from various groups around the world are not yet able to explain
some recent observational findings \citep{SchumacherSreenivasanSolConv2020} which
reveal even qualitatively
different spectrum of convection compared to what is routinely seen in most
advanced simulations \citep{Featherstone2016HighRa,kapyla2021Pr}.
The problem of stellar convection in general is quite important as it
plays a central role in governing the dynamics of
magnetic fields in stars. Moreover, it drives the seismic modes that carry useful
information about the stellar structure. As current simulations
are often understood in terms of the standard mixing length theory
\cite{bohmvitenseBookVol3}, this leads us to a serious theoretical challenge
where we seem to lack a basic understanding of convection in turbulent regimes
that are applicable to stars due to their very large Rayleigh numbers.
This provides us with an opportunity where more computing resources are expected
to lead to significant progress to understand this `convective conundrum'.

\end{itemize}

In order to make direct comparison with observations, the simulation cases listed above use significant computational resources, and involve a strong collaborative framework with large globally-distributed teams.

\begin{table*}
    \centering
    \begin{tabular}{|p{2cm}|p{4cm}|p{4cm}|p{5cm}|}
    \hline 
    \rowcolor{Gray}
    
    \textbf {Code Names} & \textbf {Equations solved} & \textbf {Application areas} & \textbf {Remarks} \\
    \hline\hline
    \rowcolor{LightCyan}
    PLUTO, FLASH, Athena++, EULAG-MHD & Hydro, MHD, some with special/general relativistic version
& Accretion, jets, feedback, turbulence, dynamos, solar corona, star-planet interaction, space weather
& Higher order Godunov methods.\\
\hline
  Gadget & Hydro with self-gravity, N-body for dark matter dynamics &
Cosmological galaxy formation & 
Smooth particle hydrodynamics\\
\hline
\rowcolor{LightCyan}
Arepo &
Hydro with self-gravity, N-body for dark matter dynamics & 
Cosmological galaxy formation &
Arbitrary Eulerian Lagrangian moving mesh code\\
\hline
PENCIL code & Hydro, MHD & Turbulence, dynamo, planet formation, accretion, 
primordial gravitational waves, 
solar MHD &
Finite difference with explicit viscosity and shock capturing\\
\hline
\rowcolor{LightCyan}
BHAC &
General Relativistic MHD &
Accretion disk dynamics around BH and compact objects & 
Finite Volume method that adopts adaptive mesh refinement\\
\hline
SPEC, SPECTRE &
GR, MHD, GRMHD (soon) &
Numerical Relativity, binary compact object dynamics &
Pseudospectral (SPEC) / Discontinuous-Galerkin (SPECTRE) discretization with relativistic shock capturing and adaptive mesh refinement. Fully nonlinear and dynamical GR.\\
\hline
\rowcolor{LightCyan}
Einstein Toolkit &
GR, MHD, GRMHD &
Numerical Relativity, binary compact object dynamics &
Finite Volume / Finite Difference methods with relativistic shock capturing and adaptive mesh refinement. Fully nonlinear and dynamical GR.\\
\hline\hline
\end{tabular}
\caption{
Some popular publicly available astrophysics codes used by Indian users. The links to these codes are as follows: PLUTO \url{https://plutocode.ph.unito.it}, FLASH \url{https://flash.rochester.edu/site/flashcode/}, Athea++ \url{https://www.athena-astro.app}, EULAG-MHD \url{https://www.astro.umontreal.ca/~paulchar/grps/eulag-mhd.html}, Gadget \url{https://wwwmpa.mpa-garching.mpg.de/gadget4/}, Arepo \url{https://arepo-code.org}, PENCIL code \url{http://pencil-code.nordita.org}, BHAC \url{https://bhac.science}, Einstein Toolkit \url{https://einsteintoolkit.org}}
\label{tab:codes_india}
\end{table*}

\subsection{Trends in Data Science, AI/ML}
 
In parallel to the explosive growth in simulated data, the availability of data from observational facilitiies has also been growing exponentially for the last several decades. This has become possible due to dramatic improvements in telescope technologies (optics, solid state detectors, etc.) coupled with even larger improvements in computer technology, largely in line with Moore’s law. Large area surveys and large numerical simulations with uniformly high-quality data and accurate and complete data provenance have become available. However, during this time of dramatic growth in the quality and quantity of astronomical data,  the number of astronomers has only been growing at a slow linear rate. In such a situation, traditional methods of data analysis on an astronomer’s desktop are no longer adequate. Large datasets need to be properly archived and advanced data products need to be produced from them. Metadata from these datasets needs to be obtained and stored in searchable databases with API access. With the advent of the World Wide Web, users expect all data to be available online through an easy-to-use web interface. These interfaces need to allow users to search through publications (e.g. using NASA/ADS), images (using DS9/Aladin), spectra and catalogs (using CADC, NED, Simbad, Vizier and Topcat). 

Astronomical data need to follow the FAIR  principle - i.e. they need to be Findable, Accessible, Interoperable and Reproducible.  To enable this, a set of common data storage formats and data exchange protocols need to be agreed upon. Fortunately, all of these are in place now through the dedicated efforts of the worldwide Virtual Observatory community, in which India has played an important role over the last two decades.

There are developments outside astronomy - particularly in the field of computer science - that are helping too. Thus data sheets (\citealt{2018arXiv180309010G}) for standardization and interoperability of metadata, and model cards (\citealt{2018arXiv181003993M}) for standardization of ML models are being widely used. These improvements in astronomical data have been accompanied by dramatic improvements in the fields of machine learning and artificial neural networks. Supported by massive investments from both public and private entities in several countries around the world, we are seeing transformational improvements in many domains. 

\begin{figure}
    \centering
    \includegraphics[width=1\columnwidth]{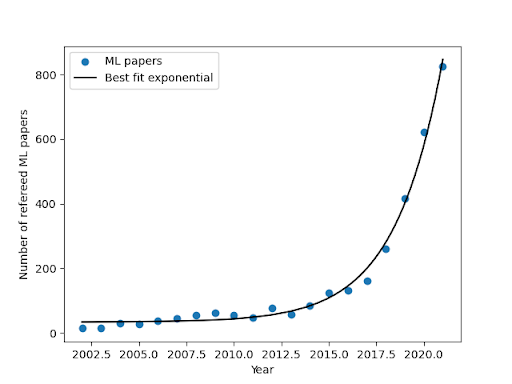}
    \caption{Number of refereed papers in astronomy listed on NASA ADS which have the phrase {\it machine learning} in the abstract in each year from 2002 to 2021. The best-fit exponential has a doubling time of just over 2 years. The 824 ML papers, written in 2021, represent about 2.5\% of the total refereed papers in astronomy in that year.}
    \label{fig:mlpapers}
\end{figure}

The application of artificial intelligence, machine learning and related techniques to almost all areas of human endeavour has seen exponential increase over the last two decades. The dramatic increase in the use of these technologies is driven by two major factors: 1. Increased computing capability  driven by dramatic improvement in hardware and 2. Widespread availability of large digital datasets for training and testing these new technologies. These two factors have caused dramatic changes in astronomy as well; as is to be expected, the use of AI/ML technologies in the astronomy domain has also increased manifold. The number of refereed papers in astronomy written the world over using machine learning techniques,  increased from 14 in 2002 to 824 in 2021 (see Figure~\ref{fig:mlpapers}). During the same period, the total number of refereed papers in astronomy showed a modest fractional increase from 19,986 to 34,811. 

There are close synergies between AI/ML and HPC; e.g., many areas like data fusion, feature extraction, classification, anomaly detection  will benefit from HPC facilities. In particular fusing data that originate with Indian facilities with data from other large surveys, and doing a combined analysis, can produce deeper scientific insights. Individuals and small groups often do that on their own. However, large and complex data volumes imply that most such groups go after low hanging fruit. Creating good datasets will result in multiple advantages: the democratisation of data usage that has been seen in the last few decades will have an added Indian flavour. The availability of high quality datasets for training will mean that a bigger software literate population can get attracted towards astronomy-based projects, creating a future generation of data-savvy astro-enthusiasts, if not astronomers.

\section{Research and status of the field in India}
\begin{figure*}
    \includegraphics[width=2\columnwidth]{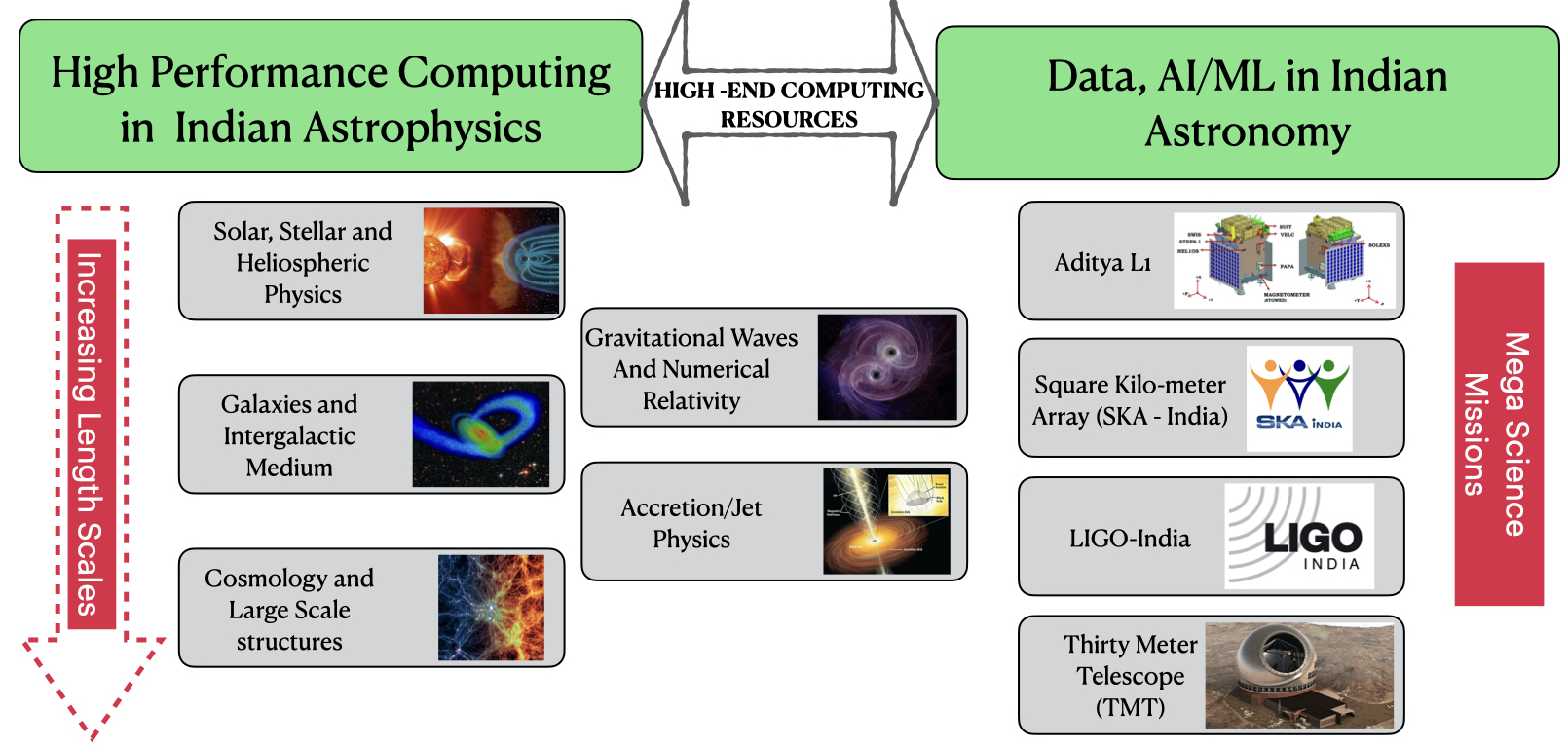}
    \caption{The current focus of Indian Astronomy and Astrophysics community in terms of application of computational astrophysics and Data science, AI/ML.}
    \label{fig:landscape_national}
\end{figure*}

In this section, we briefly review the research carried out by the various Indian A\&A groups using HPC. We also highlight some of the contributions of the Indian A\&A community in  big data and applications of AI/ML. The landscape depicting the current focus of Computational Astrophysics and Data Science, AI/ML within the Indian Astronomy and Astrophysics community is shown in figure~\ref{fig:landscape_national}.

\subsection{Computational Astrophysics in India}
India has strong groups in some areas of computational astrophysics whose contributions we highlight below according to the subdomains within A\&A.

\subsubsection{Solar, Stellar and Heliospheric Physics}
There are a number of computational modelling groups dispersed across various Indian institutes that seek to comprehend the non-linear and complex interaction of physical processes from the solar interiors to the interaction of tenuous solar wind with the Earth's magnetosphere (inner heliosphere).

Understanding the emergence of magnetic fields from inside the sun needs numerical investigation of convection, which plays a key role in the dynamo process. Few organisations in India are engaged in the 
investigation of the solar dynamo mechanism with global magneto-hydrodynamic (MHD) simulations (e.g., \cite{Karak_2015}).

As we noted above in Sec~\ref{comp_global}, 
the physics of stellar convection is not fully
understood, especially in
light of some recent observations. This is likely the reason that we do not yet
understand the origin of global and local magnetic fields in the Sun and stars, as it is
the convection that governs the evolution of magnetic fields.
Amongst the most challenging aspects of modelling the
convectively unstable layers
of stars is to adequately resolve the upper layers where the scale heights drop steeply.
Current models either do not include this layer, or use artificial schemes, such as
flux-limited diffusion, which enables the codes to run, but it is unclear if
those results can be reliable. More realistic simulations are being tried, but we are
severely limited in terms of resources that are required to make further progress.
\cite{Gopalakrishnan+23} recently proposed a simplified shell model for stratified convection where they suggest that the suppression of convection in deeper layers of the solar convection zone may be addressed without necessarily appealing to rotation or magnetic fields.

Several groups conduct full-scale 3D numerical simulations of magnetic field dynamics in the solar corona (e.g., \cite{Srivastava_2017}, \cite{Sarkar_2017}, \cite{Dey_2022}, \cite{Bora_2022}).  The purpose of this investigation is to comprehend the physical process responsible for the observed heating of the coronal plasma.
The simulations shown here necessitate the addition of kinetic physics and/or multi-fluid MHD, 
non-ideal phenomena such as Hall effect, ambipolar diffusion, thermal conduction, and other radiative effects. Therefore, these models are computationally intensive, but significant progress has been made in 
the physics of solar spicules, waves, localised plasma dynamics, and their heating capacities. In addition to evolving MHD fields, spectral synthesis codes (e.g. MULTI3D, RH, FoMo) are needed to forward model MHD simulation data in order to get emission intensities from solar chromosphere and transition region as seen by various space and ground-based detectors. These simulations and forward modelling will be essential to comprehend the data acquired from the many instruments of Aditya-L1 in space and the National Large Solar Telescope on the ground.

In addition, efforts are ongoing to 
study the properties of magnetic turbulence in the solar corona using the fluid MHD method \citep[e.g.,][]{Makhwana_2020, Sen_2021}.  Ideally, the MHD PIC (particle-in-cell) method should be used to investigate the impact of magnetic reconnection on particle acceleration, especially during solar flares. To handle the vast kinetic range of solar wind, simulations that couple PIC (kinetic physics) with MHD (continuum approach) must be developed in near future  (e.g., \citealt{Gupta2021,Paul_2021}).

Few groups in our country are 
focusing on the plasma dynamics between the Sun and the Earth, a region that extends beyond the solar corona. Murchison Widefield Array may be utilised to analyse the propagation of CME and perform computationally intensive data processing in the radio band to estimate the $z$ component 
(along earth's magnetic dipole)
of the magnetic field, which is essential for space weather research \citep[][and references therin]{Oberoi_2022}. Few groups are also working in forecasting and assessing solar wind parameters at L1 \cite{Kumar_2020, Mayank_2022} and exploring the influence of shocks on planetary magneto-spheres from a modelling perspective  \citep{Bharti_Das_2019, Basak_2021, Paul_2022}.

Numerical investigations in the area of magneto-helioseismology are now being carried out in India.
The primary goal here is to understand the plausible interaction between seismic modes and
the magnetic fields of the Sun, which supports a wide variety of
waves that carry useful information about the in-homogeneous solar
structure.
Employing the methods of helio-seismology provides
a way of looking underneath the surface of the Sun using naturally occurring sound and surface gravity ($f$) modes.
\cite{Singh_2014} carried out numerical simulations to find that the $f$-mode is significantly perturbed by near surface magnetic fields, and it fans out in the diagnostic $k-\omega$ or the dispersion diagram. Further controlled numerical experiments revealed the strengthening of the $f$-mode in presence of localized sub-surface magnetic fields \citep{Singh+20}. These studies have led to development of novel method to forecast the active region (AR) emergence using the techniques of local helio-seismology,
\cite{Singh_2016}. \cite{Singh+15} also studied the properties of acoustic and gravity modes in details using numerical simulations.  
%

\cite{Hanasoge_2008} studied wave propagation in magnetized regions and sunspots using numerical simulations with prescribed background models. This method allows for the examination of wave propagation without addressing convection and permits the use of various background models. 
More recently, several studies using helioseismology have established constraints on the upper bounds of velocity amplitudes and demonstrated that convective velocities within the Sun are much smaller than previously predicted \citep[e.g.,][and references therein]{hanasoge22}.

Further, due to complex interplay of physical processes and due to vast disparity in scales, mostly 2D simulations are possible currently to study solar convection and more computational power would be required to unravel the wave interaction within our Sun. However, with the advent of AI/ML tools, deep learning techniques can play a crucial role in the interpretation of measurements of oscillation data from various stellar sources. For example, in case of red giants and delta Scuti stars, utilizing observations from the TESS and Kepler missions. Deep-learning models have been successfully applied to infer all relevant seismic parameters from measurements of oscillation spectra for red giants \citep{dhanpal22} and some delta Scuti stars \citep{panda24}. This approach has provided new insights into the evolutionary stages of red giants \citep{dhanpal23}, inferences of their rotation rates and magnetic fields \cite{bhattacharya24_mag}. The integration of machine learning with observational data represents a major methodological advancement, enabling more detailed characterizations of stellar interiors and
their dynamic behaviors. 

\subsubsection{Gravitational Waves \& Numerical Relativity}

Gravitational-wave astronomy has a few computationally intensive aspects. On one hand we have 
to search for continuous gravitational waves with their long integration times, on the other hand the searches and follow-ups of several compact binary coalescence events are run every month. These are both collaboration-led efforts and many groups in India are already engaging constructively as part of the LIGO Scientific Collaboration. We note that the bulk of their computational needs are supported through the US NSF, with various Universities around the globe and in India contributing a fraction.
 The LIGO-India project has set up dedicated computing facilities for this purpose at nodal institutions in India, and these now await suitable and timely upgrades. We also note that all of this requires \textit{high-throughput computing}, which is quite different from and substantially less expensive than high-performance computing, which is used by most other computational projects summarized here. On the other hand, our theoretical understanding of the compact binaries we observe coalescing  in the GW spectrum has become increasingly reliant on numerical simulations due to the dynamic nature of gravity and its complex interplay with hydrodynamics, magnetic fields, and radiation transfer. 

In order to comprehend the physics in the region of BH-BH and NS-NS binaries, numerical relativity must be utilized. A number of groups in India, including ICTS, IUCAA, IISER Kolkata have initiated research into fully dynamical numerical relativistic magneto-hydrodynamics simulations of astrophysical systems. Numerical relativity simulations that evolve dynamical spacetimes can also generate synthetic gravitational wave signatures that can be compared to LIGO-Virgo data to help constrain the models. These simulations are essential for comprehending short gamma-ray bursts, stellar evolution, and the formation of binaries. However, such simulations need millions of CPU hours. Therefore, efforts have been put into using surrogate models based on high-precision numerical relativity (NR) simulations in order to restrict the physical parameters describing heavy ($\gtrsim 50 M_\odot$) black hole merging events (\citealt{Kumar_2019,Varma_2019csw}).

In addition, India is implementing a new continuous gravitational wave (CW) search pipeline and pursuing the discovery of indications of gravitational physics beyond General Relativity.
Such a pipeline will aid in the detection of weak gravitational wave signals from spinning neutron stars in binary systems with established sky placements, in addition to facilitating observations from upgraded LIGO, KAGRA, etc. (\citealt{MukherjeeA_2022}).

The current generation of numerical relativity codes capable of performing GRMHD simulation of neutron star binaries and accreting black hole binaries are both insufficiently accurate, primarily because they are unable to utilize increasing amounts of parallel computing efficiently, and therefore will be unable to translate the increased computing capacity of global supercomputing facilities of the next decade to simulation accuracy. These codes were developed by collaborative community effort spread over the US and Europe primarily, with little Indian involvement. Going forward, some groups in India have joined the global effort toward the next generation of NR codes capable of performing more efficient and accurate GRMHD simulations. Of note are the community-developed Einstein Toolkit and the SpECTRE code developed by the SXS collaboration. Going forward we will need the strong participation of the Indian community in this area so as to be part of path-breaking discoveries in the future that will inevitably come as third-generation terrestrial GW detectors and space-based LISA come online. We will also need strong backing in the form of logistical (i.e., hardware and expertise) support that this computational research will inevitably need and thrive on.

\subsubsection{Accretion and Jet Physics}

Jets are essentially collimated beams of plasma that originate from the underlying accretion disks and impact the environment at various scales. Predominately, these jets are believed to be driven via magnetic fields, and the shocks and turbulence in them play a vital role in accelerating particles. 

Numerical simulations of the magnetized accretion flow around black holes \cite{Dihingia_2021, Mishra_2022, Dhang_2022} have shown presence of turbulent inflow of matter along with a well-structured jet and disk-wind. Also, post-processing these simulations can generate synthetic images of electromagnetic radiation closer to the black hole for comparison with the Event Horizon Telescope (EHT) \cite{Dihingia_2022}.
Understanding, the origin of magnetic fields in the underlying disks has gained significant attention within India. These fields are crucial as they play a key role in the linear and nonlinear evolution of the magnetorotational instability (MRI). A consistent and more robust study of the origin of magnetic fields in the accretion disc requires an accretion flow of a large dynamical range evolved for a long integration time (e.g., \citealt{Dhang_2020}) using relativistic MHD simulations to avoid the effects of inner boundary conditions. The plasma thermodynamics also needs to be incorporated appropriately by including radiation and heating due to reconnection, etc. 

Magnetically driven jets associated with AGNs, X-ray binaries etc. are prone to instabilities that can be mediated by 
magnetic field and shear flow between the jet and the ambient medium. Large-scale 3D simulations by \citet{Mukherjee_2020, Borse_2021} have explored the role of instability in jet dynamics and have demonstrated the impact of Kelvin-Helmholtz and kink instabilities on jet flow structure. \citet{Kishor_2022} have also performed simulations to study the time-dependent relativistic jets under the influence of radiation field of the accretion disk and have explored the effects of thermodynamics on jet flow. 

Further, \cite{Vaidya_2018} developed a hybrid framework that combines Eulerian grid and Lagrangian particles to compare the dynamical jet simulations with dominant non-thermal emission and spectral signatures. 
\cite{Kundu_2021} have applied this framework to study emission signatures from AGN jets at all scales and have updated it to include particle acceleration from stochastic processes along with the diffusive shock acceleration. 
\cite{Acharya_2023} studied the role of MHD instabilities in driving multi-band variability correlations for the case of Blazars using this framework. 
Further at kpc scales, the hybrid framework have been applied to understand the formation of X- and S-shaped dual AGN jets \citep{Giri_2022b, Giri_2022} and radio lobes \cite{Kundu_2022}. 

\subsubsection{Galaxies and Intergalactic Medium}

Modern galaxy formation simulations include: (i) large-scale cosmological simulations of gravitational interactions between dark matter and baryons, and important baryonic processes such as star formation and feedback heating due to energy injection by star formation and accretion on to supermassive black holes; and (ii) idealized simulations of isolated halos, typically with much higher resolution, but which focus on limited but the most important physical elements. Since galaxy formation is highly nonlinear, analytical approaches have limitations and numerical simulations are necessary.

Recent N-body simulations have studied the growth of bars in galaxy disks and the bar-halo interaction has been shown to be an important part of disk evolution (\citealt{Kataria2018}). 
Some groups have studied bar formation in thin and thick discs \citealt{Ghosh2022}. The halo shape also has an effect on the bar evolution and disk dynamics (\citealt{Kumar2022}). Simulations of galaxy interactions are also important for understanding how disks respond to large scale gravitational forces (\citealt{Kumar2021}).

In addition to the dynamics of galaxies, groups in India are also working on the circumgalactic and intracluster medium (CGM/ICM), the diffuse extended gaseous atmospheres around galaxies. It is well-recognised that almost all galaxies in the Universe have supermassive black holes in their centres which control their growth via feedback heating (e.g., \citealt{Khandai2015}). Analogously, supernovae regulate star formation in smaller galaxies (examples of superbubble simulations done by Indian groups are \citealt{Roy2013,Yadav2017}). In addition, environmental factors like tidal/ram-pressure stripping affect galaxy properties.

Feedback heating of the ICM that roughly balances cooling losses motivated thermal instability models in which the ICM is in rough thermal balance but denser regions can cool in a runaway fashion producing cold gas.  This gas, being denser, can fall in and power feedback sources (black hole accretion and supernovae). Some of the important works in this area were carried out by Indian researchers (e.g., \citealt{Sharma2012,Choudhury2019}). Idealized driven turbulence simulations of stratified and multiphase ICM/CGM were also performed (e.g., \citealt{Mohapatra2021,Mohapatra2022}). More realistic simulations with feedback AGN jets tied with the cold gas at $\sim 1$ kpc, produced heating and cooling cycles in cool-core clusters (\citealt{Prasad2015}) and highlighted the need for fast accretion on to the black hole facilitated by the turbulent multiphase gas (e.g., \citealt{Prasad2017}).

On smaller scales, simulations of AGN jets interacting with a clumpy ISM have been carried out (\citealt{Mukherjee2016}). The jets can thermalize at small scales and can locally enhance star formation and globally suppress it. Recent observations have shown that low power radio jets are ubiquitous in galaxies and play an important role in feedback at small scales. Simulations of inclined jets interacting with a clumpy ISM have also been carried out (\citealt{Mukherjee2018}) and show that for a clumpy ISM the outflow can be symmetric despite anisotropic energy injection by jets. These questions have implications for the appearance of Fermi/eRosita bubbles in the centre of our Galaxy (e.g., see \citealt{Sarkar2022}). Indian groups have also modeled these bubbles both as supernova and AGN driven outflows (\citealt{Sarkar2015}) and constrained the power of these outflows comparing with the observations of OVIII/OVII line ratios (\citealt{Sarkar_2017}).

On even smaller scales, \citet{Kanjilal2021} have studied the growth of radiative cooling layers because of the cooling of mixed gas with a short cooling time, produced when cold/dense clouds move relative to a diffuse ambient medium. Such mixing layers may be responsible for the ubiquitous multi-temperature gas observed in quasar absorption line surveys of the CGM. Similar mixing layers have also been studied (\citealt{Dutta2022}) in clouds identified in massive halos of Illustris TNG50 simulation (\citealt{Nelson2020}).

\subsubsection{Cosmology \& Large Scale Structure}

Cosmological N-body simulations in India started with the group led by Jasjeet Bagla, who wrote a TreePM code (e.g., \citealt{Bagla2002,Bagla2005,Bagla2006,Khandai2009}). 
Recent trend has been to analyse large, easily accessible data from popular cosmological N-body/galaxy formation simulations (run and made available by large teams) to understand the complex nonlinear physics of large scale structure and galaxy formation.

\ps{Detailed studies of perturbations in dark energy at small scales require a fully relativistic approach as perturbations in matter are non-linear, and a relativistic perturbations theory cannot be used.  A 1+1 numerical relativistic approach has been used to simulate dark energy and dark matter perturbations in a self-consistent manner at small scales.  The key results of these studies has been to demonstrate that dark energy perturbations remain very small even in the non-linear regime for dark matter.  Further, it is shown that differences between different classes of dark energy models are too small for observations to discriminate between them; only the expansion history of the universe can be determined \citep{Rajvanshi2018, Rajvanshi2020, Rajvanshi2021}.}

Analyzing a large volume of cosmological galaxy formation simulation (\citealt{Khandai2015}), \citet{KarChowdhury2021}  showed that the X-ray emission in galaxies and clusters decreases in the vicinity of the black hole because of a decrease in gas density due to feedback heating. They combined cosmological simulation data with plasma emission codes and produced mock surface brightness maps using realistic {\em Chandra} telescope response. A similar approach can also be used to produce mock thermal Sunyaev-Zeldovich maps. 

In addition to X-rays, galaxy clusters also emit in radio because of plasma jets launched by central black holes and in form of diffuse radio halos/relics powered by mergers. \citet{Sur2021} have simulated fluctuation dynamos to produce polarized synchrotron emission and Faraday rotation measure maps. While hydro/MHD simulations evolve the X-ray emitting thermal plasma, the nonthermal emission from ICM/IGM has to be modeled via subgrid prescriptions; e.g.,  \citet{John2019} used diffusive shock acceleration (DSA) prescription at shocks to study cosmic ray luminosity in different stages of cluster mergers. Similar prescriptions can be used for relativistic electrons to produce radio maps (due to synchrotron emission) of large scale structure (\citealt{Paul2020}). Ultimately, we have to understand the physics of particle acceleration in shocks and turbulence from kinetic simulations using, e.g., the particle-in-cell (PIC) simulations. Prescriptions based on these first-principles simulations can be incorporated into larger scale hydro/MHD simulations. Sometimes cosmic rays are also modeled as a fluid (\citealt{Gupta2021}) with diffusion along local magnetic fields. Cosmic rays can produce a dramatic change in appearance of star clusters in different wavebands (\citealt{Gupta2018}).

Simulating the epoch of reionization using radiative transfer simulations using GPUs has been critically important in understanding the Lyman-$\alpha$ forest. These simulations produced a concordant model of reionization for the first time (\citealt{Kulkarni2019}), which led to the first robust understanding of the time at which cosmic reionization was completed. In future, such simulations will be crucial in modeling the 21-cm signal for, e.g., the SKAO. Radiative transfer to produce absorption line profiles is necessary to produce statistics such as column density distribution function of Lyman-$\alpha$ forest (e.g., \citealt{Gaikwad2017}). These statistics are necessary to constrain reionisation history and geometry, and the thermal state of the IGM (\citealt{Gaikwad2020}). 

\ps{An approach based on catastrophe theory classification has been used to develop analysis software for gravitational lensing systems.  Given a lens model, this can be used to predict the locations of all the stable and unstable caustics in the system at all possible source redshifts \cite{Meena2020}. This is of immense utility in view of upcoming large-scale surveys with Euclid and unprecedented deep observations by Roman Space Telescope and the James Webb Space Telescope (JWST).}

\begin{figure*}
    \includegraphics[width=2\columnwidth]{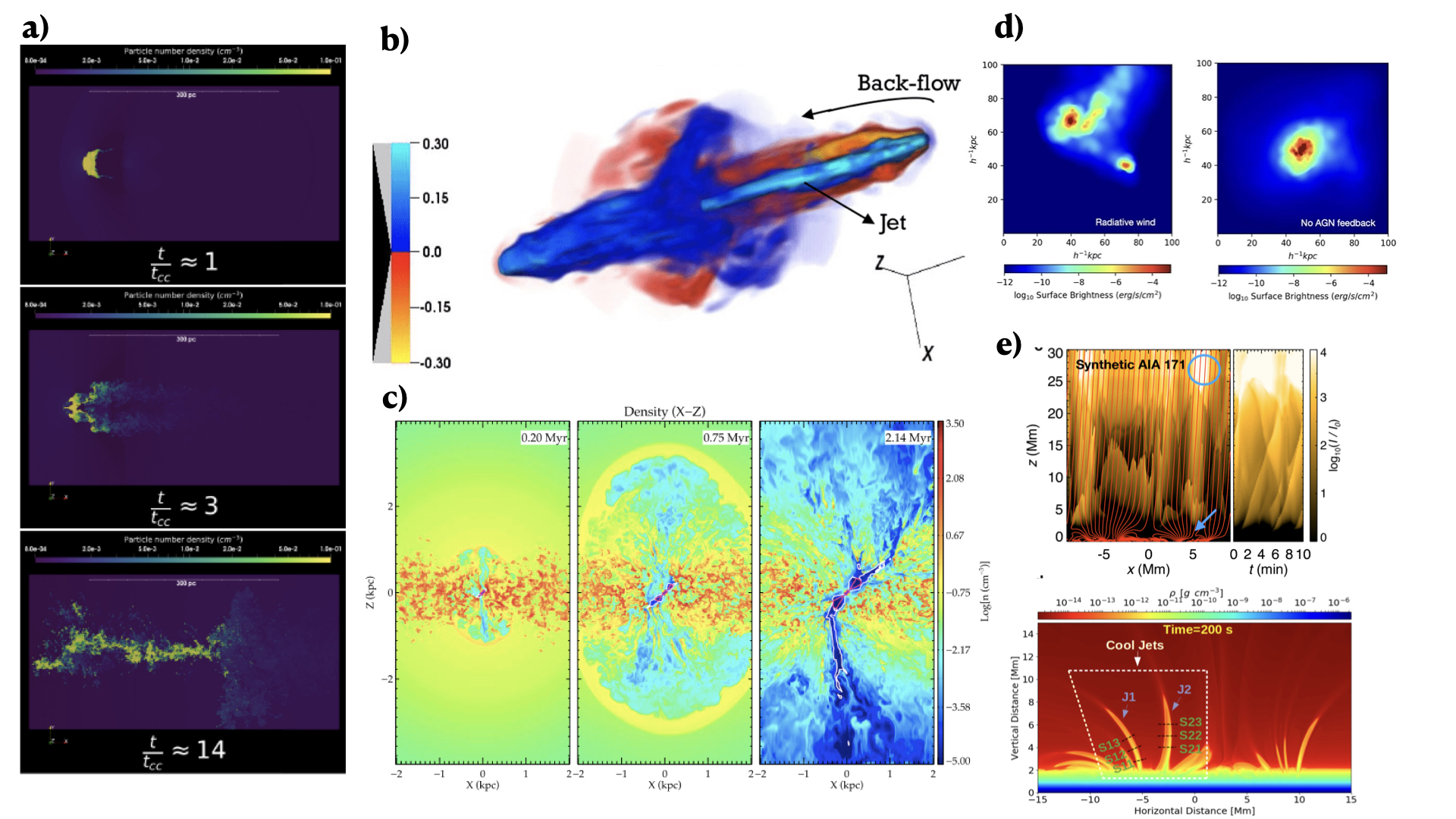}
    \caption{ 
    (\textit{Panel a}): Density snapshots from a radiative cloud-crushing simulation showing the production of mixed gas and the resulting growth of cold tail. Such radiative boundary layers may be behind the ubiquitous cold gas observed in quasar absorption lines from the circumgalactic medium (adapted from Fig. 3 of \cite{Kanjilal2021}) (\textit{Panel b}): 3D Volume rendering of jet velocity in an X-shaped radio galaxy showing the formation of winged structure due to jet back-flow (adapted from \cite{Giri_2022}. (\textit{Panel c}): Density snapshots of a simulation of inclined hydrodynamic AGN jets thermalizing on interacting with a clumpy ISM. While the inner jet is inclined, the bubble and outer shock are symmetric (adapted from Fig. 14 of \cite{Mukherjee2018}). (\textit{Panel d}) Synthetic X-ray maps generated from cosmological simulations showcasing the different modes of feedback at large scales (adapted from \cite{KarChowdhury2021}). (\textit{Panel e}) The top part shows the synthetic AIA 17.1 nm image from MHD simulations of spicules from \cite{Dey_2022} and the bottom part shows the evolution of multiple spicule-like cool
    jets in the structured solar atmosphere with realistic temperature profile from \cite{Singh_2023}.
    }\label{fig:national_hpc_sims}
\end{figure*}

A collection of some representative results from numerical simulations carried out by Indian research groups covering the above subsections is shown in figure~\ref{fig:national_hpc_sims}. 
\ps{It should be noted that the results shown in the figure (and discussed in this paper) are merely representative and do not cover the full spectrum of research work done by the Indian community.}

\subsection{Data, AI/ML in Indian Astronomy}

Large, well-organised, searchable archives of data obtained with Indian telescopes are relatively rare. The largest public archive of Indian astronomical data is the GMRT data archive\footnote{http://naps.ncra.tifr.res.in/goa}. This archive hosts all interferometric observations obtained with the GMRT since the first public cycle in 2002. As of early 2023, it contained more than 700 TB of data. It has been used by thousands of users in more than 40 countries \citep{Wadadekar2018}.

Papers applying artificial neural networks to the solution of problems in the astronomy domain began to appear in India around the beginning of the 21st century; e.g. \citet{Philip2002} used trained neural networks for the separation of stars and galaxies in optical images. Since then  there has been a dramatic increase in the use of these technologies by Indian astronomers. In 2021, as many as 47 refereed research papers on this topic included at least one co-author from an Indian institution. Several other papers were led by Indian students and postdocs working abroad.

At the present time, several research groups in India are applying AI/ML techniques to a variety of domains. We list a few representative examples on the applications of AI/ML techniques that illustrate the diverse techniques being used and the different fields of astronomical research in which they are being applied.

\begin{itemize}
    \item Transient science is now generating vast amounts of data which needs to be classified in real-time for efficient follow-up. Radio frequency interference (RFI) also produces transient artifacts which needs to be flagged and removed in real time. AI/ ML methods are very well suited for the problem of RFI identification. RFI signals have certain characteristics which are not shared by the celestial signal and experienced humans are able to visually identify them. It therefore follows that if an AI network can be trained on real-life samples of clean and RFI-ridden data, it can be used to identify RFI contaminated data.  Over the medium term,  real-time RFI mitigation in the GMRT backend systems covering the full spectrum of RFI behavior (broadband and narrowband, intermittent and persistent, etc.) is planned using AI/ML techniques

\item Algorithmic developments are also happening rapidly in both supervised and unsupervised learning domains. Both these approaches benefit from the use of a probabilistic Bayesian approach where the classification or regression value is accompanied by a probability (e.g. \citet{Abraham2012}. This approach usually requires the use of input priors. 

\item  Large area surveys are now processing prodigious quantities of data. Even basic tasks like morphological classification in the optical \citep{Grover2021} and the radio \citep{Samudre2022} now need the application of AI/ML techniques.

\item Solar imaging pipelines running on data from the MWA (Murchison Widefield Array) are now producing a large number of images (2.76 $\times  10^6$ single pol images for each hour of data). They are a ripe target for
using AI/ML algorithms for pursuing {\it computer-aided discovery} approaches as well classification approaches using training data-sets. AI/ML techniques have been used for identifying and characterising Weak Impulsive Narrowband Quiet Sun Emissions (WINQSEs) in tens of thousands of solar radio images \citep{Bawaji2023}.

\item Detection and classification of gravitational-waves from merging black-holes and neutron-stars, and the follow-up of each merger event, are seeing increasing involvement of AI/ML algorithms that can robustly reduce the latency of these tasks and enable richer multi-messenger astronomy~\citep{Khan_2021vbf,Chen_2020lzc,Wei_2020xrl,Ravichandran_2023qma}.

\item The use of AI/ML techniques is not restricted to handling real data alone. They can be used effectively with simulated datasets as well, e.g. \citet{Choudhury2022} used artificial neural networks to extract the 21-cm Power Spectrum and the
reionisation parameters from mock datasets.

\end{itemize}

\section{Key science questions, future goals, etc. from an Indian perspective}
\subsection{What do we need to do to build on our strengths?}
Some areas in computational astrophysics in which India has strong groups are: simulations of thermodynamics of intra-cluster/circumgalactic/intergalactic medium, AGN jet physics both in terms of understanding the process of particle acceleration and its feedback on multi-phase ISM, galactic dynamics, accretion around compact objects, solar dynamo and magnetoconvection, star-planet interaction and space weather modelling. 
Further, some areas like solar radio imaging, astrophysical transient detection, and radio astronomy that have adopted automation/AI/ML have put Indian groups on the world map. 
We need to go deeper/wider in these areas to produce high-impact work.

A necessary condition to become a world leader in these areas is quick access to appropriate HPC resources. Another important direction to have an impact is training of students/workforce on cutting-edge topics, and attracting best scientists/faculty from the world-wide pool of researchers. Collaborations within India and with leading groups in the world with complementary skills will help. Similarly, contributions to community codes and new code development (when open codes are not available) across areas will be important.

\subsection{Identify the emerging opportunities and how do we make their best use?}
The global hardware landscape in HPC is moving towards heterogeneous computing with GPUs becoming increasingly dominant. This is leading to the development of community codes that can efficiently utilise this hardware; the Indian community needs to be in touch with these developments to efficiently utilize the current/future HPC hardware. To be updated with the latest developments and techniques and to build a sense of community in Indian computational astrophysics, it will be important to organize periodic schools/workshops in different aspects of HPC. Private industry can also be roped in for training in and exposure to the latest tools.

In the AI/ML domain, rapid developments are taking place on many fronts like generative pretrained transformers for text generation, convolutional neural networks utilising deep learning for image classification, and generative adversarial networks for text-to-speech, text-to-image and text-to-video applications. New advances in AI/ML are being announced, almost on a daily basis. The holy grail of these efforts is artificial general intelligence (AGI), a system performing a wide range of tasks, capable of human-level intelligence, such as problem-solving, decision-making, and perception.  Enormous resources are being invested globally towards realising this goal of flexible and adaptable generalised AI. Already, rapid developments in the area of large language models (LLMs) are enhancing efficiency in writing code through tools like OpenAI's ChatGPT and GitHub's Copilot. The growing demand for efficient and powerful AI processing capabilities has driven the development of new hardware technologies such as the Neural Processing Unit (NPU). Modern NPUs support a wide range of AI models, including deep learning, reinforcement learning, and traditional machine learning algorithms. Robust software development kits (SDKs) and frameworks are being developed to facilitate the programming and deployment of AI models on NPUs. The adoption of open standards like Open Neural Network Exchange (ONNX) is promoting interoperability and ease of model transfer between different NPU platforms. We should carefully study these developments and adopt and adapt these new emerging technologies to the solution of research problems in astronomy. 
Such an approach will be particularly important in the era of Big Data from upcoming facilities like the SKAO and Rubin/LSST, where the sheer volume of data will make traditional analysis methods completely impractical.

\subsection{Weakness in country-wide portfolio and tips for improvement}

India still does not have large research groups in several areas of computational astrophysics, e.g., star/planet formation, kinetic simulations of astrophysical plasmas, and numerical relativity. The Indian universities/institutes should aggressively hire world-class faculty in these areas, some of whom can be undergraduate and graduate students trained in India working at the forefront in different leading groups across the world. We can attract excellent people only if we provide quick and fair access to HPC resources needed for them to carry on and build upon their influential research. Computational astrophysics should also spread in teaching institutes (IISc, IISERs, IITs, etc.) and universities to quickly build a large workforce required to become a leader in this important area.

Considerable expertise is available in the private industry in the area of big data and AI/ML, however, we have only been able to tap this expertise to a limited extent. More systematic efforts are needed by the Astronomy Community to harness this resource.

\subsection{Use of international facilities, collaborations, etc.}

Opportunities to access global supercomputers from India are only few and far between. Therefore, a reliable HPC infrastructure, that can serve the need of the Indian research community, must be created within the country. The National Supercomputing Mission (NSM) is a step in this direction but a framework for quick and fair access to national HPC resources still needs to be worked out.
Indian computational/data astrophysicists are collaborating with other leading groups in the world and, sometimes because of this, are able to access HPC infrastructure outside the country. However, this is no substitute for a robust, indigenous, HPC ecosystem.

\section{Facilities}
\subsection{Observational} 
High performance computing is essential not only for processing and post-processing of large volumes of astronomical data from observational facilities, but also for running simulations to interpret the physical nature of the astronomical objects 
and for determining the optimal observing and data reduction strategy to maximise the science returns of the observational facilities. Such a synergy between HPC and observational facilities is particularly important  for the success of large projects with Indian involvement such as SKAO, LIGO, TMT, Rubin/LSST and Aditya-L1. Such projects must  allocate at least 5-10\% of the total funds to HPC systems to carry out simulations and post-processing related to their scientific goals. Some large projects have already made such a projection, e.g. as part of the proposed Indian participation in the SKAO, India has proposed to construct a SKA Regional Centre within the country \citep{Wadadekar23}. This  centre will provide storage at the Exabyte scale and commensurate compute, that will be used both for SKAO data analysis and for carrying out simulations in areas related to SKAO science goals. 

\subsection{Computational}
India pursued indigenous supercomputer development programmes after being denied importing supercomputers from Cray. The latest initiative on large scale HPC in India is the national supercomputing mission (NSM\footnote{\url{https://nsmindia.in}}), a Rs. 4500 Crore initiative under which petaflop HPC resources of various architectures are being established at different research institutes/universities. About a dozen NSM clusters are already deployed at various institutes and several are in the pipeline. In addition to these supercomputers, there are smaller clusters (up to 100s of TF) at various institutes that are used in specific domains by research groups within the institutes. In February 2023\footnote{\url{https://cdac.in/index.aspx?id=pk_itn_spot1300}}, the Centre for Development of Advanced Computing (C-DAC) and the National Centre for Radio Astrophysics (NCRA) announced that they were building a high-performance computing facility with a computing capacity of 1 PetaFlop for conducting real-time commensal search for FRBs (Fast Radio Bursts) and pulsars with the GMRT. The initiative is funded by the National Supercomputing Mission (NSM) and the system will be based on the indigenous Rudra server developed by C-DAC. Once completed, this is likely to become the largest astronomy-specific NSM installation in the country.

Presently the NSM clusters at different institutes are mostly used by the community within respective institutes, with a fraction of time available for external users. The different NSM clusters at various institutes follow different policies in selection of projects and allocation of computing time. The centralized allocation of NSM cluster resources will help in a more efficient utilization and better scientific outcome. Moreover, collaborative application development projects involving PIs across institutes should be encouraged to spur innovation and to create a broad user community.

The US and European model with a central administration and proposal evaluation for all NSM machines (maintaining a fair share for the host institute) in a unified framework with information on how to access the resources at different levels will be essential to achieve optimum returns on investment in HPC. Moreover, periodic analysis of job data, similar to what is done internationally, to assess usage patterns across disciplines and applications will help fine tune the policies for maximum scientific return.

On the software front, most of the Indian astrophysics simulators are using the various open astrophysical community codes (see Table \ref{tab:codes_india}), in which they are also contributing various important modules (e.g., \citealt{Gupta2021}, \citealt{Vaidya2017}). These codes have been tested on large supercomputers worldwide for scalability and require minimal effort to get started with the research problem. 
Moreover, a lot of these codes are moving towards GPU versions to harness the power of GPU cores increasingly available at leading supercomputers.

Further, as part of this chapter we have collated data on the requirement of HPC resources from the community 
along with the way they manage their computing access now. Additionally, we have also sought information on application of AI/ML to their research in Astronomy and Astrophysics. The  
statistics from the survey are shown in figure~\ref{fig:hpc_usage_stats}.

Now we list some short and mid/long term goals for progress in HPC and AI/ML in A\&A that came out of our discussions and feedback from the community.

\subsection{Short Term Goals}
\begin{itemize}
\item Organizing regular schools/workshops for capacity building and also introducing the latest developments in the fields of HPC and data science, AI/ML.
\item Leveraging existing International Collaborations and building new ones.
\item Building an organic ecosystem of computational and data-intensive astronomy research in India by attracting the best researchers from across the world.  
\item Formulating fair policies for ready access to national super-computing facilities for competitive projects. 
\item Provisioning a separate budget head in national-level funding opportunities for computational resources, particularly when national supercomputing facilities are not readily accessible. 
\item Developing and implementing a model for industry interactions and collaboration with a pointed focus on Data Science and AI/ML in Astronomy. 
\end{itemize}

\subsection{Mid-Term/Long-Term Goals}
\begin{itemize}
\item Developing a plan for national Astronomy data centre which will act as a repository for Big Data in Astronomy both from national and international observational facilities and simulations. The center will provide resources for data analysis and develop AI/ML-based pipelines for data reduction and visualization. The centre will also provide technical support to users of the datasets hosted by the centre.
\item Developing an Inter-disciplinary Centre for Computational Research covering various fields including Astrophysics. Such a center would probably operate under a Public Private Partnership model with significant private funding. International examples of such initiatives are the Flatiron institute in New York\footnote{\url{https://www.simonsfoundation.org/flatiron/}}, and HITS in Heidelberg.\footnote{\url{https://www.h-its.org/}}
\end{itemize}

\section{Broader appeal and applicability of the field: Societal and Social Impact}
We recognize that the skills in HPC, Data Science, and AI/ML areas are in great demand across sciences and beyond. The workforce we train in HPC/data/AI/ML applied to Astronomy will have essential transferable skills that can be applied to various sectors outside of Astronomy. In the long run, this will also forge deeper connections with the industry. 
\subsection{Skill Development Programs and Capacity Building}
\begin{itemize}
    \item Expanding the coverage of HPC, Data Science, and AI/ML in Astronomy to universities and institutions without this expertise. 
    \item Creating a network of the Indian HPC community for easy sharing of expertise and knowledge transfer, fostering deeper collaborations. In this context, partnerships with computer science departments at various institutions may be mutually beneficial.
    \item Exploring possibilities to foster growth such as active collaboration/short-term visits from universities to HPC experts in A\&A domain, 
    and between astrophysicists and industry experts.
    This will be, of course, restricted by availability of supporting funds, but can be placed as an ideal solution for meaningful growth.
    \item Organising periodic workshops on basic HPC followed by focused schools on certain HPC topics. The schools can be either focused on a specific type of HPC application in A\&A (such as particle simulations, finite volume methods applied to astrophysics), or a specific architecture (such MPI based A\&A codes, GPU computing, Data/AI/ML techniques). The venues can be rotated amongst different institutes and the model evaluated and tweaked every 2-3 years to bring in fresh inductees. 
    Experts from the industry can be roped in to connect with the real-world applications of the tools that are used in astronomy research.
    \item Creating Prize Postdoctoral/Chair Faculty positions in these important areas. 
\end{itemize}

\subsection{Industrial Connect and Private Funding}

In the AI/ML domain, some astronomical institutions have already commenced working with software companies in a collaborative manner. For example, an informal collaboration between the NCRA and the Engineering for Research (E4R) group at \textit{Thoughtworks India} starting in 2019, used deep learning networks to predict star-formation related properties of galaxies using their broad-band luminosities  in the optical and infrared bands \citep{Surana20}. In 2022, TIFR initiated a much larger effort called LIBS - Laboratory for Interdisciplinary Breakthrough Science - where scientists across all campuses of TIFR in the country proposed to work with the research wing of large software companies (Google, Microsoft etc.) in a collaborative model on the application of AI/ML technology to problems in their respective science domain. TIFR scientists would provide the domain expertise in their field and the software industry partner would bring in the AI/ML expertise as well as modest annual funding to cover travel expenses to national and international conferences. 

The learnings from the early efforts such as LIBS will determine the best practices to be followed for the continued success of future undertakings. Sub-domains of astronomy that are extremely data-rich will be the most fruitful areas for industry collaboration. Large surveys in various wavebands (and their corresponding simulations) which provide large amounts of uniformly high-quality data are the first choice for the application of AI/ML techniques. Our field is fortunate to have all these data accessible through national and international survey collaborations.

\subsection{Citizen Science Programs}
The applicability of supervised machine learning techniques to astronomical problems is currently severely limited by the lack of availability of high-quality training data. Well-designed citizen science projects can be extremely useful for the generation of such training data. Steps may be taken to facilitate the development of novel citizen science projects by the Astronomy community with this aim in mind. Collaboration with existing citizen science projects  \citep{Hota2022} will also be mutually beneficial.

\section{Summary}
To summarize, HPC, big data and AI/ML have become all pervasive tools in all areas of modern astronomy and astrophysics research, and their importance is only growing with time. To be an important player worldwide, the Indian A\&A community needs to recognise the importance of these modern tools and train human resources in this fast-paced field. And perhaps most importantly, at the national level we need to quickly formulate a mechanism through which we provide serious researchers with adequate computing resources through regular calls for computational proposals that are evaluated by different domain experts and HPC specialists.


\appendix

\section*{Statistics on Usage of HPC and AI/ML}
\begin{figure}
    \includegraphics[width=0.9\columnwidth]{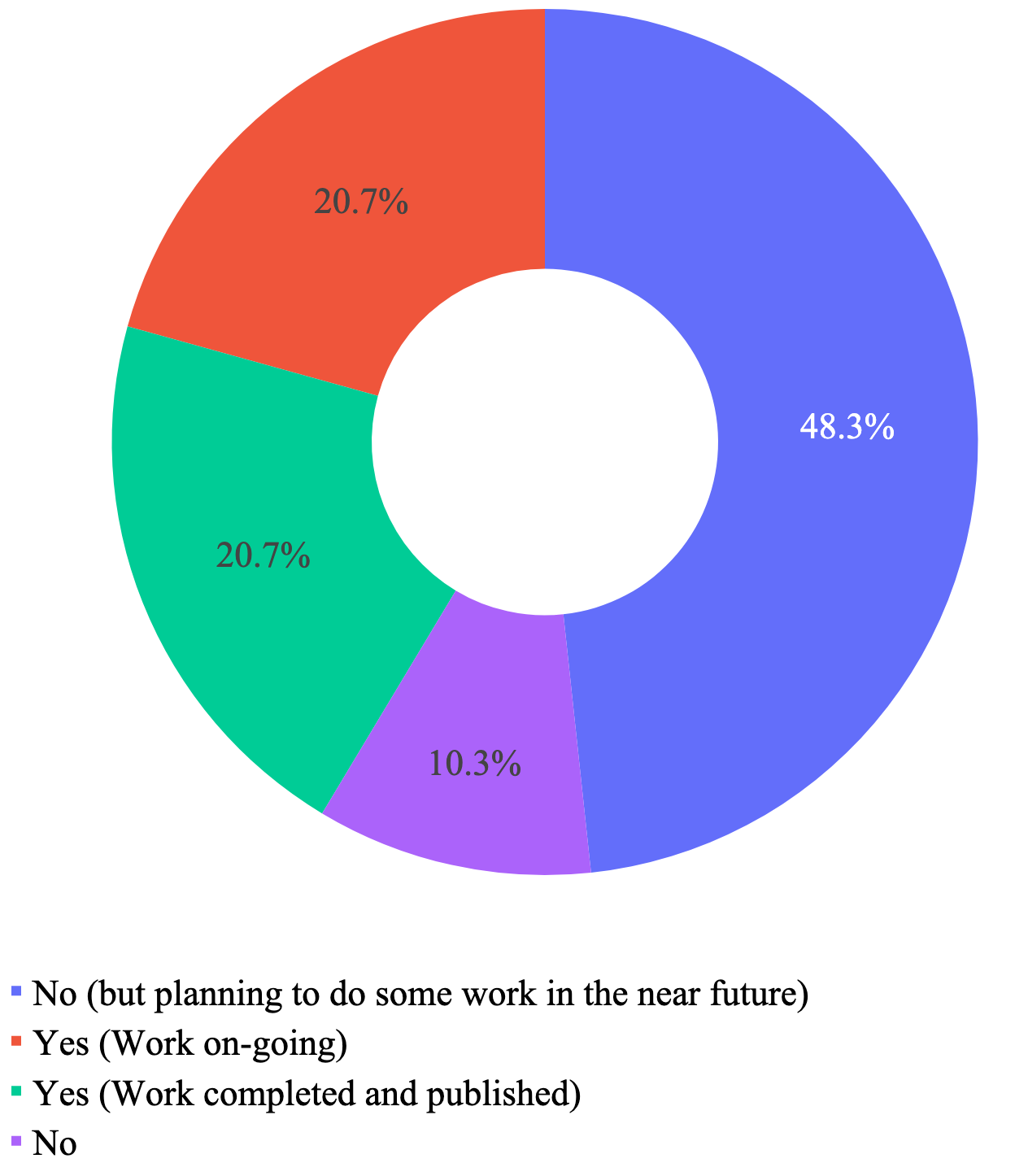}
    \caption{A pie chart indicating the application of AI/ML in main stream astronomy research. The percentages are obtained from a survey of 29 respondents.} 
    \label{fig:hpc_usage_stats}
\end{figure}

As part of the form required to provide inputs to this chapter, the contributors had three specific questions to respond with regard to HPC usage, accessibility to HPC resources, and application of AI/ML in their research. The statistical finding from about 29 responses received is shown in the figures~\ref{fig:hpc_usage_stats}. 

\begin{figure}
    \includegraphics[width=1.2\columnwidth]{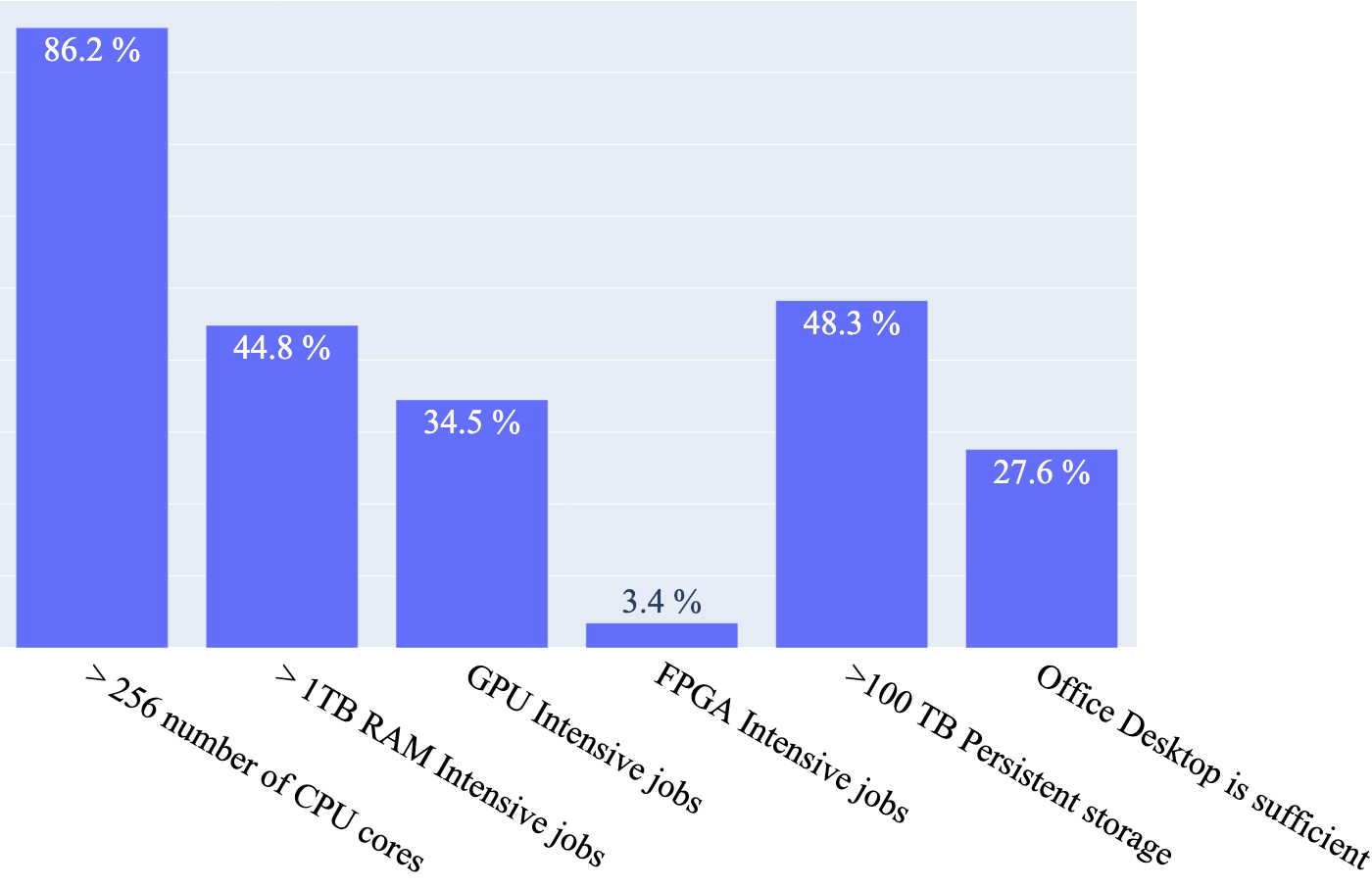}\\
    \includegraphics[width=1.2\columnwidth]{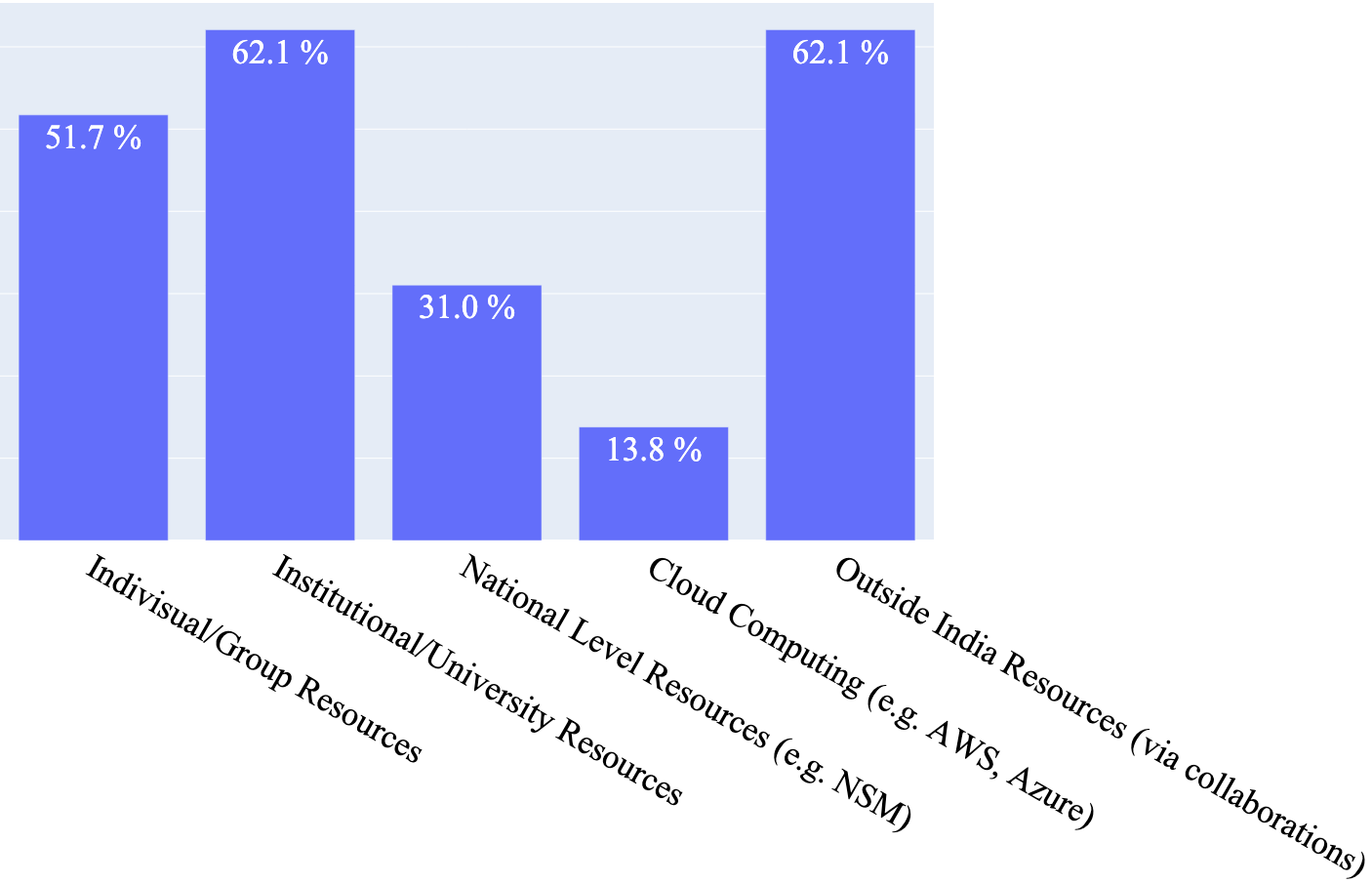}
    \caption{Bar charts showing the high performance computing usage from a survey of 29 respondents. 
    The top panel shows percentage of HPC usage by the community. The bottom panel how such computational resources are accessed by the community}
    \label{fig:hpc_usage_stats2}
\end{figure}

\section*{Contributions to the Chapter}

The following list includes the members of the Astronomical Society of India        (ASI) Vision Document 22-23 working group on \textit{Computational facilities, Techniques, Data science \& AI/ML in Astronomy}, who also appear in the author list. The list also includes contributors who responded with their suggestions through email and the online form that we circulated.

  Prateek Sharma (IISc, Bangalore, \textbf{Convenor}),
  Bhargav Vaidya (IIT Indore, \textbf{Convenor}),
  Yogesh Wadadekar (NCRA-TIFR, \textbf{Convenor}),
  Jasjeet Bagla (NCRA-TIFR (on sabbatical from IISER Mohali)),
  Varun Bhalerao (IIT Bombay),
  Ankush Bhaskar (Space Physics Laboratory, VSSC, ISRO),
  Kaushal Buch (NCRA-TIFR),
  Piyali Chatterjee (IIA),
  Suchetana Chatterjee (Presidency University, Kolkata),
  Prakriti Pal Choudhury (University of Cambridge, UK),
  Mousumi Das (IIA, Bengaluru),
  Abhirup Datta (IIT Indore),
  Prasun Dhang (University of Colarado, Boulder, USA),
  Indu Kalpa Dihingia (Tsung-Dao Lee Institute, Shanghai, China),
  Sudip K Garain (IISER Kolkatta),
  Siddhartha Gupta (University of Chicago, USA),
  Shravan Hanasoge (TIFR, Mumbai),
  Ananda Hota (UM-DAE Centre for excellence in Basic Sciences, Mumbai),
  Bidya Binay Karak (IIT BHU),
  Nishikanta Khandai (NISER, Bhubaneshwar),
  Girish Kulkarni (TIFR, Mumbai),
  Prayush Kumar (ICTS-TIFR),
  Ashish Mahabal (Caltech, USA),
  Kirit Makwana (IIT Hyderabad),
  Bhupendra Mishra (Harish-Chandra Research Institute (HRI)),
  Wageesh Mishra (IIA),
  Arunava Mukherjee (SINP, Kolkatta),
  Dipanjan Mukherjee (IUCAA),
  Divya Oberoi (NCRA-TIFR),
  Surajit Paul (Savitribai Phule Pune University, Pune),
  Ninan Sajeeth Phillip (Artificial intelligence Research and Intelligent Systems- airis4D),
  Nishant Singh (IUCAA),
  Abhishek Kumar Srivastava (IIT BHU),
  Nitin Yadav (IISER Trivendrum).

\vspace{-1em}

\bibliography{refs.bib} 

\end{document}